# Drag coefficient for the air-sea exchange in hurricane conditions

Ephim Golbraikh,[1] and Yuri M. Shtemler,[2]

[1]*Department of Physics, Ben-Gurion University of the Negev, P.O. Box 653, Beer-Sheva 84105, Israel*

[2]*Department of Mechanical Engineering, Ben-Gurion University of the Negev, P.O. Box 653, Beer-Sheva 84105, Israel*

## Abstract

A physical model is proposed for the prediction of the non-monotonic variation of the drag coefficient, $C_d$, with wind speed. The model approximates the effective $C_d$ by the area-weighted averaging of the distinct drag coefficients associated with the foam-free and foam totally-covered portions of the sea surface, and identifies the roughness of the sea surface totally covered by foam with the foam bubble size. Based on the available optical and radiometric measurements of the foam fractional coverage and the foam bubble size, the present model yields the effective $C_d$ which is in fair agreement with that obtained from field measurements of the vertical variation of mean wind speed (Powell et al., 2003), which showed a reduction of $C_d$ with wind speed rising to hurricane conditions. The present approach opens new opportunities for modeling of drag coefficient in hurricane conditions by using radiometric measurements instead of direct wind speed measurements.

## 1. Introduction

The drag reduction of the air- sea interface with wind speed rising to hurricane conditions has been a focus of intensive experimental investigation over the last decade. Many theoretical studies, laboratory and field experiments have been conducted to determine the sea-surface drag variation with wind speed in hurricane conditions (Powell et al. 2003,2007; Donelan et al., 2004; Black et al., 2007; Troitskaya et al., 2012; Soloviev et al. 2012). A reduction of the sea-surface drag in hurricane conditions instead of its monotonic growth with wind speed predicted by the Charnock relation commonly employed in moderate wind conditions (Charnock, 1955), has been found by (Powell et al. 2003). As conjectured in Powell et al. (2003), the foam cover increase due to wave breaking forms a slip surface at the atmosphere-sea interface that leads to a drag reduction at hurricane wind speeds. Saturation in drag growth has been observed in laboratory experiments by Donelan et al. (2004) who note "one may expect a qualitatively different behavior in its frictionobrezat' po vertikali skazherm do al properties than that suggested by observations in moderate wind conditions".

The principal role of the air–sea foam layer has been first suggested in (Newell and Zakharov 1992). According to empirical data, the foam formation is highly correlated with wind speed and sea gravity waves breaking (Stogryn, 1972; Monahan and O'Muircheartaigh, 1980; Monahan and Woolf, 1989; Reul and Chapron, 2003; Callaghan et al., 2007). The foam fractional coverage (foam fraction), observed after sea gravity waves break, rapidly and monotonically increases with wind speed with no saturation effects (e.g. El-Nimri et al. 2010). When $U_{10}$ exceeds the storm force (~24 m/s), wave breaking creates streaks of bubbles near the sea surface. As the wind exceeds the hurricane force (~32 m/s), streaks of bubbles combined with patches of foam cover the sea surface. When $U_{10}$ reaches ~50 m/s, a foam layer almost completely covers the sea surface (e.g. Reul and Chapron, 2003; Powell et al., 2003, El Nimri et al. 2010).

The wind speed measured at a distance above the sea surface has been extrapolated by using the log- law model of the wind profile to its zero value at the roughness height near the sea surface in order to evaluate the effective drag coefficient and sea surface roughness length in storm and hurricane conditions (Powell et al., 2003). It is to be noted that the drag coefficient estimated in Powell et al. (2003) from measurements of wind speed is naturally averaged over alternating foam-free and foam-covered portions of the sea surface. This procedure completely determines the average roughness, $Z_0$, the drag coefficient, $C_d$ and the friction velocity, $U_*$ vs. wind speed at the reference height. In turn this provides the logarithmic profile of the wind speed for theoretical modeling of the atmosphere- sea interaction in hurricane conditions (Chernyavski et al., 2011), which reduces the modeling of the sea surface stability to the effect of the log- law wind profile based upon Powell's data for the effective roughness. They also demonstrate that the wind stability model for hurricane conditions based on Charnock's formula with the standard constant coefficient underestimates the growth rate (the coefficient of the exponential growth with time of small perturbations of the air-sea interface induced by a logarithmic wind) ~5–50 times as compared with the model that employs the roughness adopted from Powell's data for hurricane winds [see the details in the above cited paper].

Another model of air-sea interaction properties in hurricane conditions was developed by (Shtemler et al., 2010). The system with the foam has been modeled by a three-fluid system of the foam layer sandwiched between the atmosphere and the sea, by distributing the foam spots homogeneously over the sea surface. They note on physical grounds that the average roughness length for the foam -atmosphere interface should correlate with the characteristic size of the sea foam bubbles. Indeed, the characteristic size of the sea foam bubbles of the order of 0.2–2mm (Leifer et al., 2003; see also Soloviev and Lukas 2006) well agree with the experimental correlation for average roughness length ∼ 0.1–2mm (Powell et al., 2003). It may be suggested that at high wind speeds the size distribution of the bubbles found on the immediate sea surface (particularly those that appear in the windrows (foam streaks), which are prevalent at high wind speeds) may be markedly different from the general bubble size distribution found in newly formed whitecaps and foam patches. Shtemler et al. (2010) demonstrate that the foam layer leads to the generation of short waves, whose length is order of roughness length of the foam-atmosphere interface, and concludes that in the three-layer model the interface roughness length is determined by the size of the foam bubbles. Their modeling exhibits a new effective mechanism of the water-surface stabilization by a foam layer, which separates the atmosphere from the sea due to high density contrasts in the three-fluid system.

In the present study a physical model is proposed for the prediction of the drag coefficient variation with wind speed as an alternative to field measurements of the vertical variation of mean wind speed by (Powell et al., 2003).

## 2. Physical model

The log-law model of the wind speed ($U$)

$$U = (U_* \ /\varkappa) \ln(Z/Z_0 ),$$ (1)

$\varkappa$=0.4 is the von Karman's constant; $Z$ is the current height over the sea surface; $Z_0$ is the sea surface roughness; $U_*$ is the friction velocity; together with the wind profile (1) the formula for the surface momentum flux, $\tau = \rho U_{*L}^2 = \rho C_{dL} U_L^2$, are commonly employed for the prediction of the drag coefficient, $C_{dL}$, variation with the neutral stability wind speed $U_L$ at a reference height $Z_L \ [m]$ ($\rho$ is the air density). This yields

$$C_{dL} = \frac{U_*^2}{U_L^2} = \left( \frac{\varkappa}{ln\left( Z_L \ /Z_0 \right)} \right)^2 .$$ (2)

Conventionally, the drag prediction problem is solved by specifying the roughness length $Z_0$.

Thus, for relatively weak winds the roughness length is well approximated by the well-known formula [Charnock, (1955)]:

$$Z_0 = \sigma_{Ch} U_*^2 / g, \tag{3}$$

where $g$ is the acceleration due to gravity, $\sigma_{Ch}$ is the phenomenological constant. In the present paper a standard value of the proportionality coefficient $\sigma_{Ch} = 0.018$ has been adopted, which provides a better correspondence of the drag coefficient with available experimental data at low winds (e.g. Large and Pond, 1981, Fairall et al., 2003, Edson et al., 2013).

For very strong winds at which the sea surface is totally covered by foam, the present model identifies the roughness length of the foam-atmosphere interface with the characteristic size of the foam bubbles, $R_b$ (Shtemler et al., 2010):

$$Z_0 = R_b . \tag{4}$$

The value $R_b$ may vary with the wind speed. However in situ measurements of foam bubble sizes are rather scarce and this dependence is not well established. In the present study we restrict ourselves to a characteristic bubble size that varies parametrically within the range of values observed for hurricane conditions (which were discussed in the introduction). The physical meaning of formula (4) becomes clear if we mentally freeze the instantaneous interface between the atmosphere and the foam layer on the sea surface. Since the average roughness of the surface is commonly defined in surface texture analysis as the mean of the absolute values of the height deviations of the surface profile, the foam-surface roughness length may be naturally identified with the characteristic radius of foam bubbles.

A standard method for determining the roughness length, $Z_0$, in a wide range of winds up to hurricane conditions yields $Z_0$ from field measurements of the vertical variation of mean wind speed at a distance from the sea surface, after extrapolation of the log- law wind profile to its zero value, $U = 0$, at the roughness height, $Z = Z_0$ (e.g. Powell et al., 2003). Since measurements of wind speed have been naturally averaged over alternating foam-free and foam-covered portions of the sea surface, this procedure provides the effective (averaged horizontally across the sea surface) values of the roughness length, friction velocity and drag coefficient, $Z_0^{(ef)}$, $U_*^{(ef)}$ and $C_d^{(ef)}$, respectively (Chernyavski et al., 2011).

In the present study, a new physical model is developed for the prediction of the drag coefficient that is also valid in a wide range of wind speeds. The proposed model treats the scaled surface momentum flux, $U_*^{(ef)2}$, as a sum of the two contributions due to additivity of the energy losses per unit surface:

$$U_*^{(ef)2} = \left(1 - \alpha_f\right) U_*^{(w)2} + \alpha_f U_*^{(f)2}, \qquad (5)$$

where $\alpha_f$ is the foam fractional coverage, $U_*^{(w)}$ and $U_*^{(f)}$ are the friction velocities for the foam-free and foam-covered sea surface areas, respectively. Correspondingly, the first term in formula (5) describes the surface momentum flux on the portion of the sea surface that is foam-free, while the second term gives the drag on that portion of the sea surface that is covered with foam patches. The value of $U_*^{(w)2}$ in Eqs. (2) is weighted by the area not covered with foam, $\left(1 - \alpha_f\right)$, since $U_*^{(w)2}$ is the proper flux only for that part of the surface. To take into account the drag reduction due to the foam covered sea surface the second term in Eq. (5) is weighted by the factor $\alpha_f$. Although the formula (5) offers the averaging procedure for the effective value of $U_*^{(ef)2}$ that differs from that proposed by (Powell et al., 2003), it may be conjectured that the linearly weighted averaging (5) yields an effective value of the friction velocity in a sense equivalent to that obtained by (Powell et al., 2003) which was based on the wind speed appropriately averaged over alternating foam-free and foam-covered portions of the sea surface.

Recognizing that the choice of the reference length cannot influence the final results, the present modeling is further tuned to the reference value $Z_{10}$, since all available experimental data are known at that altitude. For the two terms on the right side (5) the roughness is specified by Charnock's formula, and its equivalence to the foam bubble radius, for foam-free and totally-foam-covered portions of the sea surface, respectively Formula (5) yields the following relation between drag coefficients at the reference height $L$:

$$C_{dL}^{(ef)} U_L^{(ef)2} = \left(1 - \alpha_f\right) C_{dL}^{(w)} U_L^{(w)2} + \alpha_f C_{dL}^{(f)} U_L^{(f)2} \equiv const. \qquad (6)$$

Let us assume for simplicity that the effective wind speed at the 10-m altitude has the same value as for the foam-free and foam totally-covered portions of the sea surface. Then Eq. (6) is reduced to the following relation for the 10-m drag coefficients:

$$C_{d10}^{(ef)} = (1 - \alpha_f) C_{d10}^{(w)} + \alpha_f C_{d10}^{(f)}. \qquad (7)$$

It is assumed here that the Charnock formula (3) that is valid for weak winds would be provide also a fair approximation for high winds if not the presence of foam on the air-sea surface. As argued by Powell et al., 2003, the foam restricts the unrealistically large drag growth with wind speed and leads to the drag saturation for hurricane winds. With this in mind, the first term in the right hand side of Eq. (7) corresponds to an sea -surface portion completely free from the foam, for which the Charnock formula (3) for the roughness length of the atmosphere- sea interface has been adopted: $Z_0^{(w)} = \sigma_{Ch} U_*^{(w)2}/g$, with $\sigma_{Ch} = 0.018$. Then substituting $Z_0^{(w)}$ in formulas (2) yields the implicit dependences of $C_{d10}^{(w)}$ and $U_*^{(w)}$ on $U_{10}$:

$$C_{d10}^{(w)} = \left(\frac{\varkappa}{ln\left(Z_{10}/Z_0^{(w)}\right)}\right)^2, \qquad U_*^{(w)} = \sqrt{\frac{gZ_0^{(w)}}{\sigma_{Ch}}}, \qquad (8)$$

where $Z_0^{(w)}/Z_{10}$ satisfies the equation that follows from the Charnock formula

$$\frac{Z_0^{(w)}}{Z_{10}} ln^2\left(\frac{Z_0^{(w)}}{Z_{10}}\right) = \frac{\sigma_{Ch}\varkappa^2 U_{10}^2}{gZ_{10}}. \qquad (9)$$

The second term in Eq. (7) corresponds to the specific case of the portion of the sea surface entirely covered by foam. In this instance the roughness length is related to the characteristic size of the foam bubbles, $R_b^{(f)}$, by formula (3): $Z_0^{(f)} = R_b^{(f)}$. In this case formulas (1) yield for $C_{d10}^{(f)}$ and $U_*^{(f)}$:

$$C_{d10}^{(f)} = \left(\frac{\varkappa}{ln\left(Z_{10}/R_b^{(f)}\right)}\right)^2, \qquad U_*^{(f)} = \frac{\varkappa U_{10}}{ln\left(Z_{10}/R_b^{(f)}\right)}. \qquad (10)$$

The foam fractional coverage $\alpha_f$ is highly correlated to the wind speed $U_{10}$ (see Fig. 1 in El-Nimri et al. 2010 and references therein). In the present study, the foam fractional coverage is our only interest, since it represents the total foam fractional coverage (white caps and streaks). As shown by El-Nimri et al. (2010), the foam fractional coverage is very close to unity beyond the wind speed $U_{10}$ of 70 m/s, while $\alpha_f$ is certainly small at wind speeds below 18$m/s$ and reaches the order of 0.1 for the wind speed 20$m/s$. Since high wind speeds for hurricane conditions are our main interest, and the major input to the foam fraction occurs as $U_{10}$ increases above 18$m/s$, the input to the foam fraction at lower wind speed is ignored in the present approximation (as shown by oblique crosses for low wind speeds $U_{10} < 18m/s$ in Fig. 1 below). While the fraction of the sea surface covered by whitecaps or foam when the wind is less than 18÷20 m/s is certainly small, this foam coverage is significant as regards its influence on air-sea exchange which is an important component of the sea biogeochemistry (see, e.g., Vlahos and Monahan, 2009). For high wind speeds the experimental data from several sources for the total foam fractional coverage (see El-Nimri et al. (2010)) have been averaged at the fixed values of $U_{10}$ before their further treatment (experimental points for wind speeds $U_{10} > 18 m/s$ are shown with squares in Fig. 1 below). This was used in the corresponding expression $\alpha_f = 0.0007 U_{10}^2 - 0.0183 U_{10} + 0.1037$ (solid line in Fig. 1 of the present paper) obtained here by means of a least squares approximation to the experimental data.

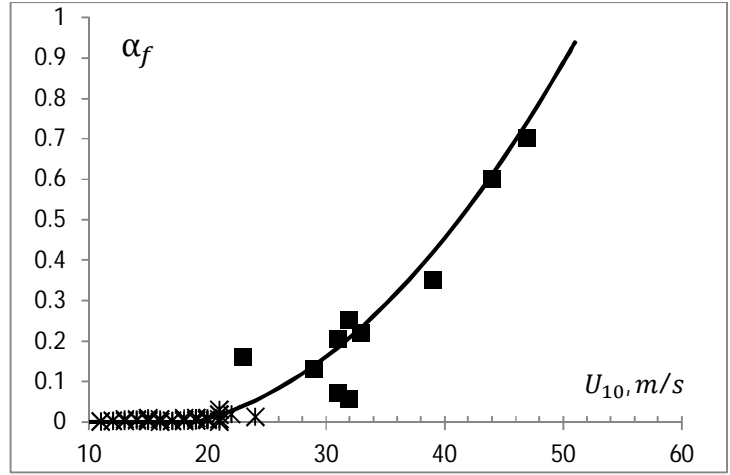

**Figure 1**. The foam fractional coverage, $\alpha_f$, vs. $U_{10}$. The least squares approximation (solid line) of experimental data for total patches (experimental data adopted from Fig. 1 in El-Nimri et al. 2010).

The resulting effective drag coefficient, $C_{d10}^{(ef)}$, which is calculated by formula (7) for several typical values of foam bubble radius, vs. $U_{10}$ along with experimental data from (Powell et al., 2003,2007; Donelan et al., 2004; Edson et al. 2006; Black et al., 2007) are presented in Fig. 2.

Then, substituting the already known $C_{d10}^{(ef)}$ in the Eqs. (2), we determine the effective roughness, $Z_0^{(ef)}$, and friction velocity, $U_*^{(ef)}$:

$$Z_0^{(ef)} = Z_{10}\exp(-\frac{\varkappa}{\sqrt{C_{d10}^{(ef)}}}), \qquad U_*^{(ef)} = U_{10}\sqrt{C_{d10}^{(ef)}}. \qquad (11)$$

The wind speed dependence of the resulting effective roughness and friction velocity are shown on Figs. 3 and 4 as well as experimental data from (Powell et al., 2003), which have been averaged for clarity over the available data sets.

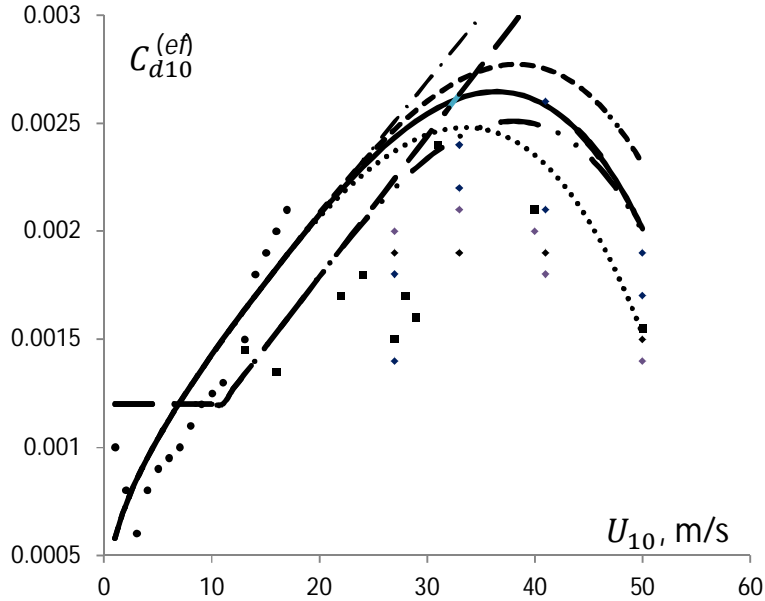

**Figure 2.** Drag coefficient $C_{d10}^{(ef)}$ vs. $U_{10}$. The small-dotted, solid and dashed curves are $C_{d10}^{(ef)}$ for $R_b^{(f)} = 0.2; 1; 2\ mm$, respectively. The dashed-dot line is $C_{d10}^{(w)}$ for the foam-free surface (Charnock's formula with $\sigma_{Ch} = 0.018$). The big-dashed line is $C_{d10}^{(w)}$ for the foam-free surface (Large and Pond, 1981). The dashed-double-dot line is $C_{d10}^{(ef)}$ (with $C_{d10}^{(w)}$ adopted from Large and Pond (1981), and $R_b^{(f)} = 1mm$). Triangles, diamonds, squares and circles are the experimental points from Donelan et al. (2004), Powell et al. (2003), Edson et al. (2007) and Black et al. (2007), respectively.

It is apparent that the resulting dependences in Figs. 2-4 for realistic values of the foam bubbles size (the main parameter of the present model) are in fair agreement with those obtained by field measurements of the vertical variation of mean wind speed.

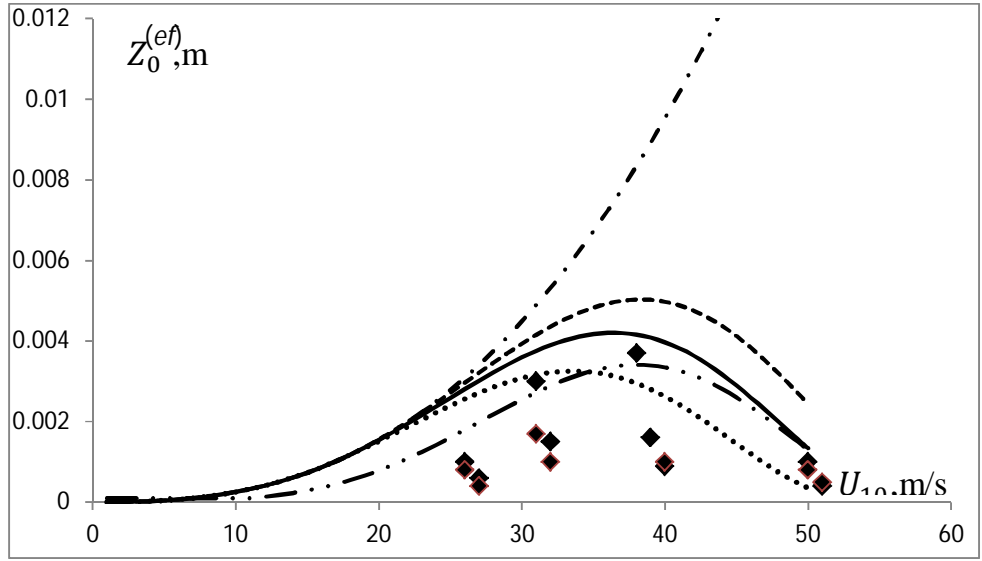

**Figure 3**. Roughness $Z_0^{(ef)}$ vs. $U_{10}$. The dotted, solid and dashed curves are $Z_0^{(ef)}$ for $R_b^{(f)} = 0.2, ; 1; 2\ mm$, respectively. The dashed-dot line is $Z_0^{(w)}$ for the foam-free surface (Charnock's formula with $\sigma_{Ch} = 0.018$). The dashed-double-dot line is $Z_0^{(f)} = R_b^{(f)}$ for the foam-covered surface (with $C_{d10}^{(w)}$ adopted from Large and Pond (1981), and $R_b^{(f)} = 1mm$). Diamonds are the experimental points adopted from (Powell et al., 2003).

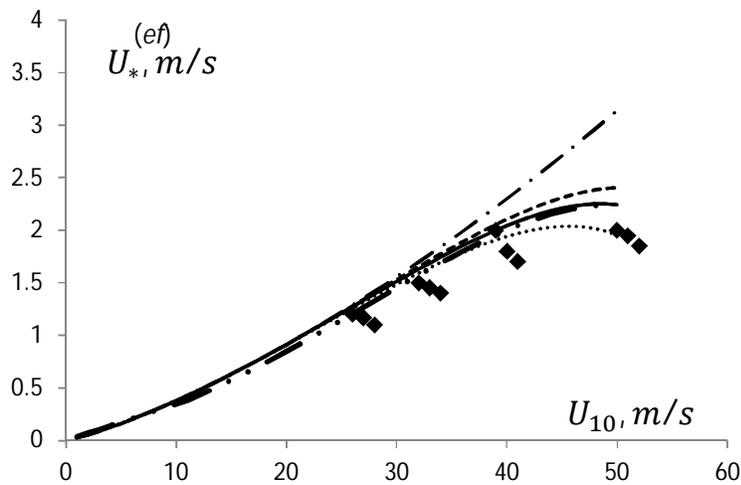

**Figure 4**. Friction velocity $U_*^{(ef)}$ vs. $U_{10}$ (the same notations as in Figure 3).

## 3. Summary and discussion

The present study is motivated by recent findings of saturation and even decrease in the drag coefficient (capping) in hurricane conditions that is accompanied by production of a foam layer on the ocean surface. The proposed model (formula (7)) treats the efficient air-sea drag coefficient, $C_{d10}^{(ef)}$, as a sum of two weighted drag coefficients, $C_{d10}^{(w)}$ and $C_{d10}^{(f)}$, for the foam-free and foam-covered conditions. As accepted in the present model, each of the three drag coefficients, one on the left side and two on the right side of Eq. (7), should obey the log law (1), but at different interface conditions for wind: real hurricane (i.e. alternating foam-free and foam-covered portions of the sea surface), foam-free and foam-covered sea-surface areas, respectively. In the present modelling of foam-free and foam-covered interface conditions, the roughness length is determined by the Charnock formula and the foam-bubble radius, respectively.

The present model is founded on the following physically based approximations:

(i) The scaled surface momentum flux $U_*^{(ef)2}$ can be presented by the area-weighted averaging of the distinct drag coefficients associated with the foam-free and foam-covered portions of the sea surface;

(ii) the effective wind speed is supposed to have the same value $U_{10}$ as for both foam-free and foam-covered conditions as those measured in real hurricane conditions, since the reference height is much larger the roughness length ($Z_{10} \gg Z_0$);

(iii) the roughness of the foam-totally-covered portions of the sea surface can be approximated by the characteristic radius of foam bubbles (the main parameter of the present model);

Although real winds over alternating foam-free and foam-covered portions of the sea surface are varied in the direction lateral to the wind propagation, any one-dimensional modeling of the vertical log profile of the wind is ultimately based on the concept of a constant (independent of the lateral coordinate) flux at the interface. This constraint is commonly alleviated by using the effective values of the drag coefficient, $C_{d10}^{(ef)}$, averaged along the lateral coordinate. The present approach based on the additivity of the energy losses per unit surface (see relation (7)) implies that the following two averaging procedures for the evaluation of $C_{d10}^{(ef)}$ are equivalent:

(i)     by measuring wind speed after it passes over alternating foam-free and foam-covered portions of the sea surface (e.g. Powell et al., 2003); and

(ii)    by averaging two model drag coefficients ( $C_{d10}^{(w)}$ and $C_{d10}^{(f)}$, each of them being constant along the lateral coordinate) with the foam fractional coverage, $\alpha_f$, as averaging coefficient.

The model describes the variations of $C_{d10}^{(ef)}$ with increases in the neutral stability 10-m wind speed $U_{10}$ from very low to hurricane force. The specific drag coefficient, $C_{d10}^{(w)}$, for the foam-free portions of the sea surface is modeled using the Charnock relation for roughness length determined by fitting the low wind data, while $C_{d10}^{(f)}$ for the foam- covered portions of the sea is modeled using the foam roughness identified with the characteristic radius of the foam bubbles. It is likely that along with $U_{10}$ other physical parameters such as atmosphere/ sea temperature difference may also affect the characteristic size of the foam bubbles. The present study applies the approximation of a constant characteristic size for foam bubbles which are parametrically varied with $U_{10}$. The characteristic size of foam bubbles has been chosen within a range of available experimental data for hurricane conditions and extrapolated down to low values of $U_{10}$, for which the influence of the deviation from its true value is reduced by low values of the weighting factor $\alpha_f$. The characteristic foam bubble size is found which provides dependence $C_{d10}^{(ef)}$ vs $U_{10}$ in fair agreement with that based upon field measurements of the vertical variation of mean wind speed. In particular, the present model describes a reduction of the drag coefficient in hurricane conditions such as was described by (Powell et al. 2003). It is seen that the maximum value of $C_{d10}^{(ef)}$ and the corresponding value $U_{10}$ are determined by the value of the characteristic size of the foam bubble.

Based on the available optical and radiometric measurements of the foam fractional coverage and of the foam bubble size, the present model yields a $C_{d10}^{(ef)}$, which is in fair agreement with that obtained from field measurements of the vertical variation of mean wind speed (Powell et al., 2003). Due to the lack of reliable data for foam bubble sizes in the whole range of hurricane conditions, in the present work the parametric runs were carried out within a realistic range of foam bubble sizes. The foam bubble sizes in that parameterization were adjusted to match our parametric modeling with the dispersion in these correlations obtained for different air-sea conditions. The adopted values of foam bubble sizes well correlate with available correlations for total drag based on the in situ measurements of the vertical velocity profile. Note that increasing of the foam-bubble radius shifts the drag maximum location to the higher wind-speed direction (Fig. 3). Variations in the foam-bubble radius in hurricane conditions may result in: either (i) monotonic growth, or (ii) saturation, or even (iii) further decrease of the drag coefficient with wind speed. In particular, a possible distinction in the foam-bubble radius may explain the observed difference in the regimes (iii) and (ii) in experiments by Donelan et al. (2004) and Powell et al. (2003) for similar hurricane conditions. For a further improvement of the proposed model, the foam bubble sizes should be measured over a wide range of wind speeds and other physical parameters in open sea conditions. The methods for optical and radiometric monitoring of the sea surface have been intensively developed in the last two decades (Amarin et al., 2012, and references therein), and these techniques combined with the method proposed here for the evaluation of the total drag can compete especially at the severest hurricane conditions with those based upon the field measurements of the velocity profile. The present approach opens new opportunities for modeling of drag coefficient in hurricane conditions by using radiometric measurements of the foam radius and foam fraction coefficient instead of direct wind speed measurements.